\documentstyle[amssymb]{article}
\begin{document}

\begin{center}
{\large {\bf Modifications of the Lorentz Force Law Complying with}}

{\large {\bf Galilean Transformations and the Local-Ether Propagation Model}}

$\ $

{\sf Ching-Chuan Su}

Department of Electrical Engineering

National Tsinghua University

Hsinchu, Taiwan
\end{center}

$\ $

\noindent {\bf Abstract} -- It is generally expected from intuition that the
electromagnetic force exerted on a charged particle should remain unchanged
when observed in different reference frames in uniform translational motion.
In the special relativity, this invariance is achieved by invoking the
Lorentz transformation of space and time. In this investigation an entirely
different interpretation of the invariance of force is presented. We propose
a new model of the electromagnetic force given in terms of the augmented
potentials, which are derived from the electric scalar potential by
incorporating a velocity difference between involved particles. The
propagation of the potentials is supposed to follow the local-ether model.
All of the position vectors, time derivatives, and velocities involved in
the proposed potentials and force law are referred specifically to their
respective frames. By virtue of this feature, the electromagnetic force is
independent of reference frame simply based on Galilean transformations. The
proposed model looks quite different from the Lorentz force law, except the
electrostatic force. However, under the common low-speed condition where the
mobile charged particles forming the current drift very slowly in a
neutralizing matrix, it is shown that the proposed model reduces to the
Lorentz force law, if the latter is observed in the matrix frame as done
tacitly in common practice.

$\ $

$\ $

\noindent {\large {\bf 1. Introduction}}

It is widely accepted that the electromagnetic force exerted on a particle
of charge $q$ and velocity ${\bf v}$ is given by the Lorentz force law in
terms of electric and magnetic fields as ${\bf F}=q({\bf E}+{\bf v}\times 
{\bf B})$. According to this famous law, the force consists of the electric
and the magnetic forces. The magnetic force is associated with magnetic
field and depends on the velocity ${\bf v}$ of the affected particle, while
the electric force is with electric field and is independent of the particle
motion. Moreover, it is well known that the current density, which generates
the magnetic field, is given by the charge density times the velocity of the
source particles, as those adopted in the formulations presented by Maxwell
[1], Heaviside [2], Lorentz [3], and Einstein [4].

It is generally expected from intuition that a force exerted on a particle
should remain unchanged when observed in different reference frames in
uniform translational motion. That is, the observed force is independent of
reference frame. Thus it was widely presumed that the velocities of the
affected and the source particles associated with the magnetic force should
have a preferred reference frame. Otherwise, the Lorentz force law is
obviously not invariant under Galilean transformations, while Galilean
invariance is common in classical mechanics, such as in Newton's laws of
motion and in the conservation laws of momentum and kinetic energy.
Historically, there were some disagreements about the reference frame of the
particle velocity among Maxwell, Thomson, Lorentz, and Einstein who
pioneered in the development of this important force law. In Maxwell's
treatise on electricity and magnetism, the velocity seems to mean the
relative velocity, since it was noticed that it is the relative motion
between one coil and another that determines the induction in Faraday's
experiments [1]. And the electric current is attributed to be associated
with ``the velocity with which the electricity passes through the body''. In
analyzing the experiments of magnetic deflection with cathode rays, which
led to the discovery of electron, Thomson supposed that the electromagnetic
force depends both on the relative velocities and the {\it actual}
velocities of the involved particles, where the actual velocity is referred
to the magnetic medium through which the particle is moving [5]. In
Lorentz's theory of electrons, the velocity seems to be referred to the
ether, since it is supposed that the charges move through the ether, which
occupies all space between particles and pervades the particles [3].

However, in the past century, an almost unanimously accepted explanation is
provided by Einstein's special relativity whereby the velocity is referred
to an arbitrary inertial frame. When observed in another reference frame,
the relevant electromagnetic quantities of charge density, current density,
potentials, and fields will change with the relative velocity between the
two frames according to the Lorentz transformation of space and time.
Meanwhile, the kinematical quantities of velocity, momentum, and force will
also change under the Lorentz transformation. It can be shown that the
transformation of the time rate of change of kinematical momentum of a
charged particle is exactly identical to the Lorentz force given in terms of
the transformed fields and velocity [6]. In other words, the Lorentz force
law is invariant under the Lorentz transformation.

While the Lorentz transformation solves the problem associated with the
changes of particle velocities and current density with reference frame, it
is noticed that the electric current generating the magnetic force in the
magnetic deflection or in motors is commonly electrically neutralized. That
is, the mobile source particles which form the current are actually embedded
in a matrix and the ions which constitute the matrix tend to electrically
neutralize the mobile particles, such as electrons in a metal wire. It is
expected that the ions as well as the mobile source particles form an
electric current and contribute to a magnetic field if the matrix is moving.
Thus the velocity which determines the neutralized current density is the
drift velocity of the mobile particles with respect to the neutralizing
matrix, regardless of the motion of the matrix. The neutralized current
density then remains unchanged when observed in different reference frames.
Further, it is noticed that the mobile source particles commonly drift very
slowly with respect to the matrix. For conduction electrons in a metal, the
drift speed is normally lower than 1 mm/sec. Thereby, if the magnetic force
is associated with the squared velocity difference between the affected and
the source particles, then, by virtue of the compensation due to the
neutralizing matrix, the magnetic force can be associated with the
velocities of the affected and the source particles with respect to the
matrix. This situation is somewhat similar to the relation $({\bf v}-{\bf v}%
_{s})^{2}-({\bf v}-{\bf v}_{m})^{2}$ $\simeq -2({\bf v}-{\bf v}_{m}){\bf %
\cdot }({\bf v}_{s}-{\bf v}_{m})$ if ${\bf v}_{s}\simeq {\bf v}_{m}$. Thus
the magnetic force as well as the neutralized current density can be
frame-independent simply based on Galilean transformations.

Recently, we have presented a local-ether model of wave propagation whereby
electromagnetic wave is supposed to propagate via a medium like the ether
[7]. However, the ether is not universal. Specifically, it is proposed that
in the region under sufficient influence of the gravity due to the Earth,
the Sun, or another celestial body, there forms a local ether which in turn
moves with the gravitational potential of the respective body. Thus, as well
as earth's gravitational potential, the earth local ether for earthbound
propagation is stationary in an ECI (earth-centered inertial) frame.
Consequently, earthbound wave phenomena can depend on earth's rotation but
are entirely independent of earth's orbital motion around the Sun or
whatever. Meanwhile, the sun local ether for interplanetary propagation is
stationary in a heliocentric inertial frame. This local-ether model has been
adopted to account for the effects of earth's motions in a wide variety of
propagation phenomena, particularly the Sagnac correction in GPS (global
positioning system), the time comparison via intercontinental microwave
link, and the echo time in interplanetary radar. Moreover, as examined
within the present accuracy, the local-ether model is still in accord with
the Michelson-Morley experiment which is known to make the classical ether
notion obsolete. Further, by modifying the speed of light in a gravitational
potential, this simple propagation model leads to the deflection of light by
the Sun and the increment in the interplanetary radar echo time which are
important phenomena supporting the general theory of relativity [7].

As a consequence, the propagation mechanism of the potentials associated
with the electromagnetic fields and force is expected to follow the
local-ether model. Thus the potentials propagate away from the location of
source isotropically at the speed of light $c$ with respect to an ECI frame
for earthbound phenomena. This propagation model also implies that the
position vectors of the source and the affected particles involved in the
potentials should be referred specifically to the associated local-ether
frame in calculating the propagation delay time in terms of the isotropic
propagation speed $c$. A further consequence is that this restriction on
reference frame necessitates the construction of a new model of
electromagnetic force which complies with the frame-independence of magnetic
force, Galilean transformations, and with the local-ether propagation model,
and is still in accord quantitatively with the well-studied phenomena known
to follow the Lorentz force law.

Accordingly, we propose in this investigation the augmented potential, which
is derived from the electric scalar potential by additionally incorporating
a squared velocity difference between involved particles. Then, in terms of
the spatial and time derivatives of the augmented potential, we propose a
new model of electromagnetic force. Each of the position vectors, time
derivatives, propagation velocity, particle velocities, and current density
involved in the proposed theory will be imposed with a deliberately chosen
reference frame. By virtue of this simple feature, the augmented potential
and the electromagnetic force are independent of reference frame, as
expected intuitively. Thereby, an entirely different explanation of the
invariance of electromagnetic force is presented. The proposed force law in
terms of the local-ether potentials looks quite different from the Lorentz
force law. However, it will be shown elaborately that in spite of the
dissimilarity and the restriction on reference frame, this new classical
model can reduce to a form quite similar to the famous Lorentz force law
under some ordinary conditions.

$\ $

\noindent {\large {\bf 2. Lorentz Force Law in Lorenz Potentials}}

The famous Lorentz force law for a charged particle can be written in terms
of the Lorenz retarded potentials. That is, 
$$
{\bf F}({\bf r},t)=q\left\{ -\nabla \Phi ({\bf r},t)-\frac{\partial }{%
\partial t}{\bf A}({\bf r},t)+{\bf v}\times \nabla \times {\bf A}({\bf r}%
,t)\right\} ,\eqno
(1) 
$$
where the scalar potential $\Phi $ and the vector potential ${\bf A}$ in
turn are given explicitly in terms of the charge density $\rho $ and the
corresponding current density ${\bf J}$ respectively as 
$$
\Phi ({\bf r},t)=\frac{1}{\epsilon _{0}}\int \frac{\rho ({\bf r}^{\prime
},t-R/c)}{4\pi R}dv^{\prime }\eqno
(2) 
$$
and 
$$
{\bf A}({\bf r},t)=\frac{1}{\epsilon _{0}c^{2}}\int \frac{{\bf J}({\bf r}%
^{\prime },t-R/c)}{4\pi R}dv^{\prime },\eqno
(3) 
$$
where $R=|{\bf r}-{\bf r}^{\prime }|$, $R/c$ represents the propagation
delay time, and $c$ is the speed of light. These retarded potentials were
first introduced by L. V. Lorenz (rather than H. A. Lorentz) in as early as
1867 [8]. It is known that the current density ${\bf J}$ is related to the
charge density $\rho $ as 
$$
{\bf J}({\bf r},t)=\rho ({\bf r},t){\bf v}_{c},\eqno
(4) 
$$
where ${\bf v}_{c}$ is the velocity of the mobile charged particles forming
the current. In terms of the Lorenz retarded potentials $\Phi $ and ${\bf A}$%
, the electric field intensity ${\bf E}$ and the magnetic flux density ${\bf %
B}$ can be given explicitly by 
$$
{\bf E}({\bf r},t)=-\nabla \Phi ({\bf r},t)-\frac{\partial }{\partial t}{\bf %
A}({\bf r},t)\eqno
(5) 
$$
and 
$$
{\bf B}({\bf r},t)=\nabla \times {\bf A}({\bf r},t).\eqno
(6) 
$$
The divergences and the curls of fields ${\bf E}$ and ${\bf B}$ given above
have been derived, which are just Maxwell's equations [9].

It is noted that in the preceding formulas, three velocities and one time
derivative which can depend on reference frame are involved. That is, the
particle velocity ${\bf v}$ connecting to the curl of potential ${\bf A}$,
the particle velocity ${\bf v}_{c}$ associated with the current density $%
{\bf J}$, the propagation velocity associated with the delay time $R/c$, and
the time derivative applied to ${\bf A}$. In the special relativity, time
derivatives as well as velocities are referred to an arbitrary inertial
frame. And the wave equation, the continuity equation, and the Lorenz gauge
which are important equations involving time derivative can be shown to be
Lorentz invariant [6].

On the other hand, based on Galilean transformations, the time derivative $%
\partial /\partial t$ observed in the chosen frame can be expressed in terms
of the derivative $(\partial /\partial t)_{k}$ observed in another reference
frame which moves at a velocity ${\bf v}_{k}$ with respect to the chosen
frame. It is known that the two time derivatives are related to each other
as [10] 
$$
\left( \frac{\partial f}{\partial t}\right) _{k}=\frac{\partial f}{\partial t%
}+{\bf v}_{k}\cdot \nabla f,\eqno
(7) 
$$
where the time derivatives $\partial /\partial t$ and $(\partial /\partial
t)_{k}$ are understood to be taken under constant ${\bf r}$ and (${\bf r}-%
{\bf v}_{k}t$), respectively, as ${\bf r}$ is referred to the chosen frame.
As the proposed force law is supposed to comply with Galilean
transformations and the local-ether propagation model, the reference frame
of each of the position vectors, velocities, and time derivatives
encountered hereafter will be given specifically.

$\ $

\noindent {\large {\bf 3. Augmented Potentials and Electromagnetic Force Law}%
}

Now we present the new electromagnetic force law given in terms of a single
scalar potential $\breve{\Phi}$, which is a modification of the electric
scalar potential by additionally incorporating a velocity-dependent part and
hence is called the augmented scalar potential. Consider an ensemble of
mobile source particles of volume charge density $\rho _{v}$. The various
source particles contribute collectively to the potential which in turn will
propagate to and then affect another charged particle under consideration
called the {\it effector}.

It is proposed that the {\it augmented scalar potential} $\breve{\Phi}$
experienced by an effector particle is given explicitly by 
$$
\breve{\Phi}({\bf r},t)=\frac{1}{4\pi \epsilon _{0}}\int \left( 1+\frac{%
v_{es}^{2}}{2c^{2}}\right) \frac{\rho _{v}({\bf r}^{\prime },t-R/c)}{R}%
dv^{\prime },\eqno
(8) 
$$
where $v_{es}=|{\bf v}_{es}|$, the velocity difference ${\bf v}_{es}={\bf v}%
_{e}-{\bf v}_{s}$, ${\bf v}_{e}$ and ${\bf r}$ are respectively the velocity
and position vector of the effector particle at the instant $t$, ${\bf v}%
_{s} $ and ${\bf r}^{\prime }$ are those of the source particle at an
earlier instant $t^{\prime }=t-R/c$, $R/c$ is the propagation delay time
from the source point ${\bf r}^{\prime }$ to the field point ${\bf r}$, and
the propagation range $R$ ($=|{\bf r}-{\bf r}^{\prime }|$) denotes the
distance from the source at the instant $t^{\prime }$ to the effector at the
field point at the instant $t$. Note that ${\bf v}_{es}$ denotes the
difference between the velocities of the effector and the source at two
distinct instants. For quasi-static case where the propagation delay time $%
R/c$ can be neglected or for the case where velocities ${\bf v}_{e}$ and $%
{\bf v}_{s}$ are fixed, the non-simultaneous velocity difference ${\bf v}%
_{es}$ then becomes the Newtonian relative velocity between the effector and
the source. It is supposed that speeds $v_{e}$, $v_{s}$, and hence $v_{es}$
are much lower than $c$, as they are ordinarily. Thus the augmentation in
the scalar potential is of the second order of normalized speed (with
respect to $c$) and is actually quite small in magnitude. It is noted that
the augmented scalar potential depends on the effector velocity and tends to
be slightly different for different effectors.

Next, we address ourselves to the issue of the reference frames of the
position vectors, the particle velocities, and the propagation velocity.
According to the evidence supporting the local-ether model, the propagation
range for an earthbound electromagnetic wave is referred specifically to an
ECI frame, if the propagation delay time is understood to be the propagation
range $R$ divided by the constant speed $c$. Accordingly, the position
vectors ${\bf r}$ and ${\bf r}^{\prime }$, which determine the propagation
range, are also referred specifically to that frame. Thus, when observed in
the associated local-ether frame, the potential propagates isotropically at
the speed $c$ away from the emission position, regardless of the motion of
the source and effector. Moreover, in order to comply with the reference
frame of the corresponding position vectors, the effector velocity ${\bf v}%
_{e}$ and the source velocity ${\bf v}_{s}$ are also referred specifically
to the local-ether frame. Thereby, all the position vectors and velocities
involved in the proposed augmented scalar potential are referred uniquely to
the local-ether frame.

Under the influence of the augmented scalar potential $\breve{\Phi}$
originating from an ensemble of mobile source particles, the electromagnetic
force exerted on an effector particle of charge $q$ and located at ${\bf r}$
at the instant $t$ is hereby postulated to be 
$$
{\bf F}({\bf r},t)=q\left\{ -\nabla \breve{\Phi}({\bf r},t)+\left( \frac{%
\partial }{\partial t}\sum\limits_{i}\hat{i}\frac{\partial }{\partial v_{ei}}%
\breve{\Phi}({\bf r},t)\right) _{e}\right\} ,\eqno
(9) 
$$
where $v_{ei}={\bf v}_{e}\cdot \hat{i}$, $\hat{i}$ is a unit vector, index $%
i=x,y,z$, and the time derivative $(\partial /\partial t)_{e}$ is referred
specifically to the effector frame with respect to which the effector of
velocity ${\bf v}_{e}$ is stationary. Physically, the time derivative $%
(\partial /\partial t)_{e}$ represents the time rate of change in some
quantity experienced by the effector. It is noticed that the proposed force
law resembles Lagrange's equations in classical mechanics adopted by Thomson
in exploring magnetic force [5].

The proposed electromagnetic force law can be rewritten in the form of 
$$
{\bf F}({\bf r},t)=q\left\{ -\nabla \breve{\Phi}({\bf r},t)-\left( \frac{%
\partial }{\partial t}{\bf \breve{A}}({\bf r},t)\right) _{e}\right\} ,\eqno
(10) 
$$
where the {\it augmented vector potential} ${\bf \breve{A}}$ experienced by
the effector is then defined explicitly in terms of the charge density $\rho
_{v}$ as 
$$
{\bf \breve{A}}({\bf r},t)=\frac{-1}{4\pi \epsilon _{0}c^{2}}\int \frac{{\bf %
v}_{es}\rho _{v}({\bf r}^{\prime },t-R/c)}{R}dv^{\prime }.\eqno
(11) 
$$
It is seen that the potential ${\bf \breve{A}}$ is associated with minus the
derivative of potential $\breve{\Phi}$ with respect to the effector speed $%
v_{e}$. It is noted that both of the augmented potentials depend on the
velocity difference ${\bf v}_{es}$ and hence tend to be different for
different effectors. As all the position vectors, the particle velocities,
the propagation velocity, and the time derivative are referred specifically
to their respective reference frame, the values of the local-ether augmented
potentials and their derivatives are indeed frame-independent. Moreover, the
corresponding electromagnetic force remains unchanged in different frames,
as expected intuitively.

Recently, based on the local-ether model, we\ have presented a wave equation
which incorporates the electric scalar potential and an operator associated
with the velocity difference. From this local-ether wave equation, the
proposed force law together with the augmented potentials and the inertial
mass of the effector can be derived from a quantum-mechanical approach [11,
12]. Anyway, in this investigation, formulas (8)--(11) serve as our
fundamental postulates for the electromagnetic force law complying with
Galilean transformations and the local-ether model of wave propagation. This
local-ether electromagnetic force law made its debut in [13].

Note that in the proposed model the force term associated with the major
part of $\nabla \breve{\Phi}$ is independent of particle velocities, while
the other terms depend on them. The velocity-dependent force terms are
expected to be much weaker than the velocity-independent one by nature,
since they incorporate the factor of $1/c^{2}$. The velocity-independent
term is identical to the conventional electrostatic force when the
propagation delay in the potentials is not considered. However, the
velocity-dependent force terms look quite different from the corresponding
ones in the Lorentz force law. The similarity will emerge under the common
low-speed condition where the source particles drift very slowly in a
neutralizing matrix, as discussed in the following sections.

$\ $

\noindent {\large {\bf 4. Augmented Potentials under Neutralization}}

Consider the common case where the ensemble of mobile charged particles are
embedded in a matrix. There are various types of the matrix, such as metal
in antenna, semiconductor in light-emitting diode, dielectric in lens, or
magnet in motor. Ordinarily, the ions which constitute the matrix tend to
electrically neutralize the mobile charged particles. Without the
neutralizing matrix, the electrostatic force will be overwhelmingly dominant
over the velocity-dependent ones. Thus the velocity-dependent forces are
observable only under sufficient neutralization. Anyway, the charge density $%
\rho _{m}$ of the matrix is supposed to be arbitrary for generality. The
neutralizing matrix can be not really in existence, for which case $\rho
_{m}=0$.

Ordinarily, the matrix is fixed in shape and moves as a whole at a
substantially fixed velocity ${\bf v}_{m}$ with respect to the local-ether
frame. Thus the various ions forming the matrix also move at this velocity.
Then, due to the ensemble of mobile source particles and the matrix, the
total electromagnetic force exerted on an effector is given by
superposition. A little algebra leads to that the augmented potentials under
neutralization take the forms of 
$$
\breve{\Phi}({\bf r},t)=\Phi ({\bf r},t)\left( 1+\frac{v_{em}^{2}}{2c^{2}}%
\right) -{\bf v}_{em}\cdot {\bf A}({\bf r},t)+\frac{1}{2\epsilon _{0}c^{2}}%
\int \frac{v_{sm}^{2}\rho _{v}({\bf r}^{\prime },t-R/c)}{4\pi R}dv^{\prime }%
\eqno
(12) 
$$
and 
$$
{\bf \breve{A}}({\bf r},t)={\bf A}({\bf r},t)-\frac{1}{c^{2}}{\bf v}%
_{em}\Phi ({\bf r},t),\eqno
(13) 
$$
where ${\bf v}_{em}={\bf v}_{e}-{\bf v}_{m}$ and ${\bf v}_{sm}={\bf v}_{s}-%
{\bf v}_{m}$. The electric scalar potential $\Phi $ and the magnetic vector
potential ${\bf A}$ in turn are redefined respectively as 
$$
\Phi ({\bf r},t)=\frac{1}{\epsilon _{0}}\int \frac{\rho _{n}({\bf r}^{\prime
},t-R/c)}{4\pi R}dv^{\prime }\eqno
(14) 
$$
and 
$$
{\bf A}({\bf r},t)=\frac{1}{\epsilon _{0}c^{2}}\int \frac{{\bf J}_{n}({\bf r}%
^{\prime },t-R/c)}{4\pi R}dv^{\prime },\eqno
(15) 
$$
where the potential $\Phi $ is due to the net charge density $\rho _{n}=\rho
_{v}+\rho _{m}$ and the potential ${\bf A}$ is due to the {\it neutralized
current density} ${\bf J}_{n}$ given by 
$$
{\bf J}_{n}({\bf r},t)={\bf v}_{sm}\rho _{v}({\bf r},t).\eqno
(16) 
$$
It is noted that the velocity of the mobile source particles forming the
neutralized current density is referred to the matrix frame with respect to
which the matrix is stationary. The velocity ${\bf v}_{sm}$ is a Newtonian
relative velocity representing the drift velocity of the source particles
with respect to the matrix. Thus the neutralized current density is
independent of reference frame. Moreover, the values of the local-ether
potentials $\Phi $ and ${\bf A}$ redefined in the preceding formulas are
also frame-independent. It is noted that unlike the augmented potentials $%
\breve{\Phi}$ and ${\bf \breve{A}}$, the potentials $\Phi $ and ${\bf A}$ do
not depend on the effector velocity and hence their distributions apply to
different effectors. Thus it can be more convenient to express
electromagnetic force in terms of $\Phi $ and ${\bf A}$ when possible.

$\ $

\noindent {\large {\bf 5. Force Law under Low-Speed Condition}}

Ordinarily, the mobile source particles drift very slowly with respect to
the matrix. Besides, the speeds of terrestrial particles with respect to an
ECI frame are normally of the order of the linear speed due to earth's
rotation, which in turn is much lower than $c$. Thus the various particle
speeds are supposed to satisfy the {\it low-speed condition} given by 
$$
v_{sm}\lll c\ \ {\rm and}\ \ v_{em},v_{m}\ll c.\eqno
(17) 
$$
In words, this ordinary condition states that the effector, sources, and
matrix move somewhat slowly with respect to the local-ether frame and the
sources drift very slowly with respect to the matrix frame. Thereby, the
augmented scalar potential becomes a simpler form of 
$$
\breve{\Phi}({\bf r},t)=\Phi ({\bf r},t)-{\bf v}_{em}\cdot {\bf A}({\bf r}%
,t).\eqno
(18) 
$$
It is noted that under the low-speed condition, the augmented scalar
potential as well as the augmented vector potential can be determined from $%
\Phi $, ${\bf A}$, and ${\bf v}_{em}$. It is owing to the drift speed $%
v_{sm} $ being very low that the matrix frame becomes the most convenient
reference frame in dealing with the electromagnetic force.

The forces due to the velocity-dependent parts of the potentials are much
weaker than the electrostatic force by nature and hence will substantially
emerge only under sufficient neutralization. That is, $\rho _{v}$ and $\rho
_{m}$ are of opposite signs and nearly the same amount, and hence $|\rho
_{n}|\ll |\rho _{v}|$. Under complete neutralization $\rho _{n}=0$, the
augmented vector potential ${\bf \breve{A}}$ becomes identical to the
potential ${\bf A}$. Suppose for the moment that the density $\rho _{n}$ in $%
{\bf \breve{A}}$ can be neglected. Then the augmented vector potential
reduces to the magnetic vector potential 
$$
{\bf \breve{A}}({\bf r},t)={\bf A}({\bf r},t).\eqno
(19) 
$$
The roles of the omitted parts in the preceding two relations of the
potentials will be discussed in a later section.

Furthermore, from the Galilean transformation (7), the time derivative $%
(\partial $${\bf A}/\partial t)_{e}$ referred to the effector frame can be
given in terms of the time derivative $(\partial $${\bf A}/\partial t)_{m}$
referred to the matrix frame. That is, 
$$
\left( \frac{\partial }{\partial t}{\bf A}({\bf r},t)\right) _{e}=\left( 
\frac{\partial }{\partial t}{\bf A}({\bf r},t)\right) _{m}+{\bf v}_{em}\cdot
\nabla {\bf A}({\bf r},t).\eqno
(20) 
$$
Thereby, the proposed force law (10) based on the augmented potentials can
be expressed in terms of the local-ether potentials $\Phi $ and ${\bf A}$ as 
$$
{\bf F}({\bf r},t)=q\left\{ -\nabla \Phi ({\bf r},t)+\nabla [{\bf v}%
_{em}\cdot {\bf A}({\bf r},t)]-\left( \frac{\partial }{\partial t}{\bf A}(%
{\bf r},t)\right) _{m}-{\bf v}_{em}\cdot \nabla {\bf A}({\bf r},t)\right\} .%
\eqno
(21) 
$$
It is noted that there are two force terms connected both with the relative
velocity ${\bf v}_{em}$ and the potential ${\bf A}$. Then, by using a vector
identity, the electromagnetic force law becomes a more familiar form of 
$$
{\bf F}({\bf r},t)=q\left\{ -\nabla \Phi ({\bf r},t)-\left( \frac{\partial }{%
\partial t}{\bf A}({\bf r},t)\right) _{m}+{\bf v}_{em}\times \nabla \times 
{\bf A}({\bf r},t)\right\} .\eqno
(22) 
$$
According to the dependence of force terms on the effector velocity, one is
led to express the force law in terms of fields ${\bf E}$ and ${\bf B}$ in
the form of 
$$
{\bf F}({\bf r},t)=q\left\{ {\bf E}({\bf r},t)+{\bf v}_{em}\times {\bf B}(%
{\bf r},t)\right\} ,\eqno
(23) 
$$
where the electric and magnetic fields are then redefined explicitly in
terms of the local-ether potentials $\Phi $ and ${\bf A}$ as 
$$
{\bf E}({\bf r},t)=-\nabla \Phi ({\bf r},t)-\left( \frac{\partial }{\partial
t}{\bf A}({\bf r},t)\right) _{m}\eqno
(24) 
$$
and 
$$
{\bf B}({\bf r},t)=\nabla \times {\bf A}({\bf r},t).\eqno
(25) 
$$
The electromagnetic force law (22) or (23) under the ordinary low-speed
condition represents modifications of the Lorentz force law, which comply
with Galilean transformations and the local-ether propagation model. The
fundamental modifications are that the current density generating the
potential ${\bf A}$, the time derivative applied to ${\bf A}$ in the
electric induction force, and the effector velocity connecting to $\nabla
\times {\bf A}$ in the magnetic force are all referred specifically to the
matrix frame and that the propagation velocity of the potentials is referred
specifically to the local-ether frame. Thereupon, the values of the current
density, the potentials, the fields, and of the force are independent of
reference frame. The divergence and the curl relations of electric and
magnetic fields together with the local-ether wave equations of the
potentials and fields are discussed in [14].

$\ $

\noindent {\large {\bf 6. Sagnac Effect and Galilean Relativity}}

In many cases it can be more convenient to deal with the propagation of the
potentials in a laboratory frame (such as the matrix frame), instead of the
local-ether frame. We then go on to consider the consequence due to the
motion of the laboratory frame. According to the local-ether model, the
propagation range $R$ in the definitions of the potentials is the distance
from the source point ${\bf r}^{\prime }$ at which the potentials are
emitted to the field point ${\bf r}$ at which the potentials propagate to
the effector, where both ${\bf r}^{\prime }$ and ${\bf r}$ are referred
uniquely to the local-ether frame. A quantity closely related to the
propagation range $R$ is the propagation-path length $R_{t}$, which
represents the geometric distance from the source to the effector right at
the instant of wave emission. The difference between the propagation range $R
$ and the path length $R_{t}$ is known as the Sagnac effect, which is due to
the movement of the effector with respect to the local-ether frame during
wave propagation. To the first order of normalized speed, it is known that $%
R=R_{t}(1+\hat{R}_{t}\cdot {\bf v}_{e}/c)$, where the unit vector $\hat{R}%
_{t}$ points from the source to the effector at the instant of emission [7].
When the wave propagation is observed in a laboratory frame moving at a
velocity ${\bf v}_{0}$ with respect to the local-ether frame, the apparent
propagation range $R_{l}$ observed in the laboratory frame is not actually
equal to the propagation range $R$. It is seen that $R-R_{l}=R_{t}\hat{R}%
_{t}\cdot {\bf v}_{0}/c$. This difference represents the deviation in the
Sagnac effect due to the motion of the laboratory frame with respect to the
local-ether frame during wave propagation. The corresponding fractional
deviation in the propagation range and time is at most of the order of the
normalized speed $v_{0}/c$, which is mainly due to earth's rotation and is
as small as about $10^{-6}$ for ordinary terrestrial laboratory frames. If
this small term of the normalized laboratory speed can be neglected, then
the apparent propagation range is substantially identical to the propagation
range. Thus, when formulated in a laboratory frame with the corresponding
Sagnac-effect deviation being omitted, the propagation range $R$, the
propagation time $R/c$, and hence the velocity differences ${\bf v}_{es}$
and ${\bf v}_{em}$ (which are associated with $R/c$ in general) become
independent of the laboratory velocity.

Therefore, the augmented potentials $\breve{\Phi}$ and ${\bf \breve{A}}$,
the local-ether potentials $\Phi $ and ${\bf A}$, the local-ether fields $%
{\bf E}$ and ${\bf B}$, and the electromagnetic force law formulated in a
laboratory frame also become independent of the laboratory velocity and
hence comply with Galilean relativity, when the Sagnac-effect deviation due
to the motion of the laboratory frame is omitted. As an example, consider
the familiar case of a cathode-ray tube, where an electron beam is
accelerated electrostatically by applying a voltage to a pair of electrode
plates and then the beam direction is adjusted by using a magnetic
deflection yoke to form an image on a screen. The yoke is a current-carrying
winding which results in a magnetic field and hence a magnetic force exerted
on the electron beam. Thus the yoke forms the matrix of this case. According
to the local-ether electromagnetic force law, the magnetic force is
associated with the velocity of electrons with respect to the laboratory
frame in which the matrix together with other components of the CRT is
stationary.

As the Sagnac-effect deviation associated with the laboratory velocity is
neglected, the magnetic field and the magnetic force in a geostationary CRT
are independent of earth's rotation and latitude of the device, let alone
earth's orbital motion. Further, when the CRT is put on a vehicle moving
smoothly with respect to the ground, the field and the force are also
independent of the ground velocity of the vehicle. Thus operation of the CRT
complies with Galilean relativity. This invariance of magnetic deflection
keeps the image on a moving CRT unaffected, which is in accord with common
experience with the CRT. A similar argument of the invariance of magnetic
force can be applied to a motor, where the current in the stator winding or
magnet results in a magnetic force exerted on the current-carrying rotor.

Meanwhile, in the conventional analysis of the magnetic deflection with the
Lorentz force law, the electron speed is commonly given by the square root
of the accelerating voltage, aside from a scaling factor associated with the
charge and mass. By so doing, the adopted reference frame is actually the
one with respect to which the electrode plates is stationary. In practical
designs, the plates are commonly stationary with respect to the deflection
yoke, although this restriction is not necessary in principle. By virtue of
this seemingly trivial coincidence, the matrix frame has been adopted
tacitly in the conventional practice dealing with the effector speed.
Besides, as the current in the yoke is neutralized, the drift speed
associated with this current is actually referred to the matrix frame
tacitly in common practice. Therefore, based on the local-ether
electromagnetic force law (22) or (23) with the Sagnac-effect deviation
being omitted, {\bf the proposed Galilean model is identical to the Lorentz
force law, if the latter is observed in the matrix frame} as done tacitly in
common practice. In other words, the Lorentz force law is only valid under
hidden restrictions. That is, the effector and the matrix speeds are low,
the drift speed is very low, the laboratory frame is actually the matrix
frame, and the corresponding Sagnac-effect deviation is omitted. However,
these conditions are so common as to be ignored easily.

Further, if the propagation delay time $R/c$ and then the Sagnac effect due
to the effector movement can be neglected, the term $1/R$ in the augmented
potentials reduces to $1/R_{t}$. Moreover, the velocity difference ${\bf v}%
_{es}$ becomes Newtonian relative and then, as well as $R_{t}$, possesses
symmetry between a pair of source and effector. Consequently, the
electromagnetic force between two charged particles given directly from (10)
complies with Newton's third law of motion: action is equal to reaction in
magnitude.

$\ $

\noindent {\large {\bf 7. Extra Force Terms}}

To complete the proposed force law, we proceed to estimate the magnitudes of
the extra forces which are omitted in the low-speed condition. From the
general forms of the augmented potentials (12) and (13), the electromagnetic
force law is given by 
$$
{\bf F}=q\left\{ -\nabla \Phi -\left( \frac{\partial }{\partial t}{\bf A}%
\right) _{m}+{\bf v}_{em}\times \nabla \times {\bf A}\right\} +{\bf F}_{1}+%
{\bf F}_{2}+{\bf F}_{3},\eqno
(26) 
$$
where the extra force terms{\it \ } 
$$
{\bf F}_{1}=-\frac{q}{c^{2}}\frac{1}{2}{\bf v}_{em}^{2}\nabla \Phi ,\eqno
(27) 
$$
$$
{\bf F}_{2}=-\frac{q}{c^{2}}\frac{1}{2\epsilon _{0}}\nabla \int \frac{%
v_{sm}^{2}\rho _{v}}{4\pi R}dv^{\prime },\eqno
(28) 
$$
and 
$$
{\bf F}_{3}=\frac{q}{c^{2}}\left( \frac{\partial }{\partial t}{\bf v}%
_{em}\Phi \right) _{e}.\eqno
(29) 
$$
All of these extra terms are velocity-dependent and incorporate the factor
of $1/c^{2}$.

The extra force terms ${\bf F}_{1}$ and ${\bf F}_{3}$ are associated both
with the effector velocity ${\bf v}_{em}$ and the scalar potential $\Phi $.
These two terms will vanish either if $\Phi =0$ or if $v_{em}=0$. In the
cases where only the magnetic force is involved, such as in the
aforementioned deflection yoke or stator magnet, the complete neutralization 
$\Phi =0$ holds. Meanwhile, in the cases where the effector is bounded to
the matrix, such as the conduction electron or the polarization charge in
antennas, waveguides, or scattering, the stationary effector $v_{em}=0$
holds substantially. Anyway, whenever the extra force term ${\bf F}_{1}$
does not vanish, it will be much weaker than the accompanying electrostatic
force ${\bf F}_{es}$ by a factor of as small as $(v_{em}/c)^{2}$ in
magnitude, which in turn is much less than unity ordinarily. That is, 
$$
\frac{|{\bf F}_{1}|}{|{\bf F}_{es}|}\sim \frac{v_{em}^{2}}{c^{2}}\ll 1.%
\eqno
(30)
$$
In other words, even when the neutralization is not sufficient, the extra
force ${\bf F}_{1}$ will be relatively too weak to observe.

In addition to the situation of complete neutralization or stationary
effector, the extra force term ${\bf F}_{3}$ can vanish in a static
potential. Even for the case where $\Phi \neq 0$, ${\bf v}_{em}\neq 0$, and
their time derivatives also do not vanish, such as in the communication
between mobile antennas, the extra force ${\bf F}_{3}$ is also negligible as
compared to the accompanying electrostatic force. It is noted that $|q\Phi
|/m_{0}c^{2}\ll 1$ ordinarily, where $m_{0}$ is the mass of the effector.
Besides, for a potential even rapidly varying with time (say, of the form of
a plane wave $\Phi (x-ct)$ or a spherical wave $\Phi (r-ct)/r$), the order
of magnitude of its time derivative divided by $c$ is not expected to exceed
the one of its gradient. Physically, this is owing to wave motion whereby a
temporal variation in a potential tends to cause a spatial variation. Thus $%
|(\partial \Phi /c\partial t)_{e}|\lesssim |\nabla \Phi |$ and hence the
extra force ${\bf F}_{3}$ will be much weaker than the accompanying
electrostatic force at least by a factor of $v_{em}/c$. That is, 
$$
\frac{|{\bf F}_{3}|}{|{\bf F}_{es}|}\lesssim \frac{v_{em}}{c}\ll 1.\eqno
(31) 
$$
Thus, even when the neutralization is not sufficient and the term ${\bf v}%
_{em}\Phi /c^{2}$ in the augmented vector potential is not negligible, the
extra force ${\bf F}_{3}$ will be relatively too weak to observe. Again,
this is because that the accompanying electrostatic force will
overwhelmingly dominate, so long as the low-speed condition holds. Thereby,
the sufficient neutralization that leads to the approximation (19) is not
necessary and the validity of the force law (23) is independent of the
extent of neutralization.

The extra force term ${\bf F}_{2}$ is due to a current and emerges even in
the situation of complete neutralization, stationary effector, and static
potential. Thus an effector stationary with respect to a loop carrying a
direct current will experience this extra force. However, it may be
difficult to observe. Like the electrostatic force this extra force is
conservative. Thus, for an effector bounded in a conducting wire, the
induced electromotive force around a closed wire is zero. On the other hand,
an unbounded effector tends to have a speed $v_{em}\gg v_{sm}$, due to
thermal fluctuation or whatever. Note that the drift speed $v_{sm}$ of the
mobile source particles with respect to the matrix is very low ordinarily.
Even for a copper wire carrying a current of density of as high as $10$
amp/mm$^{2}$, the drift speed of the conduction electrons is merely as low
as $7.4\times 10^{-4}$ m/sec. Thus the very low drift speed ($v_{sm}\lll c$)
holds ordinarily. Then the extra force ${\bf F}_{2}$ is ordinarily much
weaker than the accompanying magnetic force ${\bf F}_{m}$ by a factor of $%
v_{sm}/v_{em}$. That is, 
$$
\frac{|{\bf F}_{2}|}{|{\bf F}_{m}|}\sim \frac{v_{sm}}{v_{em}}\ll 1.\eqno
(32)
$$
Even in the case where the effector speed $v_{em}$ happens to be quite low
and hence the preceding relation does not hold, the extra force ${\bf F}_{2}$
as well as ${\bf F}_{m}$ will be too weak to observe. In the worst case
where the drift speed $v_{sm}$ is no longer quite low (such as in a plasma),
the neutralization tends to be difficult to maintain and then the
electrostatic force dominates again. Consequently, all the three extra force
terms can be neglected as compared to the accompanying electrostatic or
magnetic force, under the ordinary low-speed condition where the effector,
sources, and matrix move somewhat slowly with respect to the local-ether
frame and the sources drift very slowly with respect to the matrix frame.

Although the electromagnetic force law proposed in this investigation is
merely for low-speed particles, it may be extended to the case where the
effector speed is high. To make an attempt for this, we tentatively assume
the force law (26) remains unaltered and then examine the extra force terms
in some high-energy experiments. For this purpose we consider the familiar
cyclotron experiment, where the high-energy particles are accelerated by an
electrostatic force and are deflected by a magnetic force to move
circularly. As the effector speed $v_{em}$ is high, the extra force ${\bf F}%
_{2}$ is relatively weak. The relation among the magnetic field, the radius
of circulation, and the particle momentum can be measured with a high
precision. It has been demonstrated that when the dependence of mass on
particle speed is taken into account, the radius is linearly proportional to
the momentum, even when the particle speed is quite close to $c$ [15]. (The
local-ether theory is in accord with the speed-dependent mass, since this
mass together with its effect on the quantum state energies in atoms has
been derived from an associated wave equation [11, 14].) Correspondingly,
the local-ether force law is in accord with the magnetic force when the
effector speed is thus high. Then the restriction on the effector speed can
be removed for the case of complete neutralization where the electrostatic
force is not involved. On the other hand, the extra forces ${\bf F}_{1}$ and 
${\bf F}_{3}$ could present a problem. However, it is noted that the
acceleration in the cyclotron is actually implemented by making the particle
enter the field due to a lower voltage repeatedly by a large number of
passages. To our knowledge, no direct measurements of the acceleration of a
high-energy particle in each individual passage are reported. Thus the
presented force law is not in immediate conflict with the cyclotron
experiment or the like. Anyway, as the speed-dependent mass has been found
to affect the influence of the electric scalar potential on quantum
energies, the electrostatic force for a high-speed particle deserves a
further study experimentally and theoretically.

$\ $

\noindent {\large {\bf 8. Conclusion}}

In order to comply with Galilean transformations and the local-ether model
of wave propagation, a new classical model of electromagnetic force is
presented in terms of the augmented potentials, which are derived from the
electric scalar potential by incorporating the velocity difference between
the effector and source. The propagation of the potentials is proposed to
follow the local-ether model and hence the propagation range is referred
specifically to an ECI frame for earthbound phenomena. Thus the position
vectors and hence the velocities of the effector and the sources are
referred to this frame. As a consequence, the values of the augmented
potentials are independent of reference frame. Furthermore, in the proposed
force law, the time derivative of the augmented vector potential is
associated with the change in the potential experienced by the effector and
hence is referred to the effector frame. Consequently, the electromagnetic
force exerted on an effector is frame-independent, as expected intuitively.
The dominant part of the force based on the augmented potentials does not
depend on particle velocities and is identical to the conventional
electrostatic force, as the propagation delay time is not considered. On the
other hand, the velocity-dependent force terms look quite different from the
corresponding ones in the Lorentz force law.

However, a situation commonly encountered in the magnetic force is that the
mobile source particles are embedded in a neutralizing matrix. Then the
augmented potentials are expressed in terms of the local-ether potentials
which are associated with the net charge density and the neutralized current
density. Further, under the ordinary low-speed condition where the effector,
sources, and matrix move slowly with respect to the local-ether frame and
the sources drift very slowly in the neutralizing matrix, the proposed force
law reduces to a form similar to the Lorentz force law, aside from three
extra force terms. This represents modifications of the Lorentz force law
which comply with the frame-independence of force, Galilean transformations,
and with the local-ether propagation model. The fundamental modifications
are that the current density generating the vector potential, the time
derivative applied to this potential in the electric induction force, and
the effector velocity connecting to the curl of this potential in the
magnetic force are all referred specifically to the matrix frame and that
the propagation velocity of the potentials is referred to the local-ether
frame. It is shown quantitatively that all these extra terms are much weaker
than the accompanying electrostatic or magnetic force under the low-speed
condition. Then the proposed Galilean model is identical to the Lorentz
force law, if the latter is observed in the matrix frame and the
Sagnac-effect deviation due to the motion of this frame is neglected.
Thereby, hidden restrictions for the Lorentz force law imposed tacitly on
particle speeds and reference frame are unveiled.

$\ $

\newpage 

\noindent {\large {\bf References}}

\begin{itemize}
\item[{\lbrack 1]}]  J.C. Maxwell, {\it A Treatise on Electricity and
Magnetism} (Oxford, {New York}, 1998, 3rd. ed. of 1891), Arts. 295, 530, and
607.

\item[{\lbrack 2]}]  O. Heaviside, {\it Philos. Mag}. {\bf 27}, 324 (1889).

\item[{\lbrack 3]}]  H.A. Lorentz, {\it The Theory of Electrons} {(Dover,
New York, 1952, 2nd. ed. of 1915), Arts. 7 and 8 and Note 2.}

\item[{\lbrack 4]}]  {A. Einstein, in {\it The Principle of Relativity}
(Dover, New York, 1952), p. 37.}

\item[{\lbrack 5]}]  J.J. Thomson, {{\it Philos. Mag}. {\bf 11}, 229 (1881).}

\item[{\lbrack 6]}]  {{P. Lorrain and D.R. Corson, {\it Electromagnetic
Fields and Waves} (Freeman, {San Francisco, }1972), chs. 5 and 6; }J.D.
Jackson, {\it Classical Electrodynamics} (Wiley, New York, 1975), ch. 11.}

\item[{\lbrack 7]}]  C.C. Su, {\it Eur. Phys. J. C} {\bf 21}, 701 (2001); 
{\it Europhys. Lett}. {\bf 56}, 170 (2001).

\item[{\lbrack 8]}]  E. Wittaker, {\it A History of the Theories of Aether
and Electricity} (Amer. Inst. Phys., 1987), vol. I, ch. VIII.

\item[{\lbrack 9]}]  C.C. Su, {\it Eur. J. Phys}. {\bf 22}, L5 (2001).

\item[{\lbrack 10]}]  See, for example, L.E. Ballentine, {\it Quantum
Mechanics} (Prentice-Hall, Englewood Cliffs, 1990), sect. 4-3.

\item[{\lbrack 11]}]  C.C. Su, {\it Eur. Phys. J. B} {\bf 24}, 231 (2001).

\item[{\lbrack 12]}]  C.C. Su, {\it J. Electromagnetic Waves Applicat.} {\bf %
16}, 1275 (2002).

\item[{\lbrack 13]}]  C.C. Su, in {{\it IEEE Antennas Propagat. Soc. Int}. 
{\it Symp}. {\it Dig.}} (2001), vol. 1, p. 208; in{\ {\it Bull. Am. Phys.
Soc.}} (Mar. 2001){\it ,} p. 1144{.}

\item[{\lbrack 14]}]  C.C. Su, {\it Quantum Electromagnetics}
(http://qem.ee.nthu.edu.tw).

\item[{\lbrack 15]}]  See, for example, M. Alonso and E.J. Finn, {\it Physics%
} (Wesley, New York, 1992), ch. 22.
\end{itemize}

\end{document}